%% file: masetti_Ke4.tex
\documentclass[12pt,english]{article}
\pdfoutput=1
\usepackage{graphicx}
\usepackage{xspace}
\usepackage{babel,varioref}

\input econfmacros.tex

\newcommand{\dk}{\ensuremath{K_{e4}}\xspace}

\newcommand{\dkext}{\ensuremath{K^{\pm} \to \pi^{+} \pi^{-} e^{\pm} \nu_e}\xspace}

\newcommand{\wssel}{\ensuremath{\pi^{\pm} \pi^{\pm} e^{\mp} \nu_e}\xspace}

\newcommand{\kthreepic}{\ensuremath{K^{\pm} \to \pi^{+} \pi^{-} \pi^{\pm}}\xspace}

\newcommand{\kthreepin}{\ensuremath{K^{\pm} \to \pi^0 \pi^0 \pi^{\pm}}\xspace}

\newcommand{\pipi}{\ensuremath{\pi\pi}\xspace}
\newcommand{\azz}{\ensuremath{a_0^0}\xspace}
\newcommand{\azd}{\ensuremath{a_0^2}\xspace}
\newcommand{\qq}{\ensuremath{\left<0\left|\bar{q}q\right|0\right>}\xspace}
\newcommand{\dkextnn}{\ensuremath{K^{\pm} \to \pi^{0} \pi^{0} e^{\pm} \nu_e}\xspace}
\newcommand{\dknn}{\ensuremath{K_{e4}^{00}}\xspace}
\newcommand{\dkpm}{\ensuremath{K_{e4}^{+-}}\xspace}
\newcommand{\mnnsq}{\ensuremath{M_{00}^2}\xspace}
\newcommand{\mpsq}{\ensuremath{m_{\pi^+}^2}\xspace}

\def\Title#1{\begin{center} {\Large {\bf #1} } \end{center}}

\begin{document}

\Title{\dk decays and Wigner cusp}

\begin{center}{\large \bf Contribution to the proceedings of HQL06,\\
Munich, October 16th-20th 2006}\end{center}

\bigskip\bigskip


\begin{raggedright}  

{\it Lucia Masetti\footnote{Present address: Physikalisches Institut, Universit\"at Bonn, D-53012 Bonn, GERMANY}\index{Masetti, L.}\\
Institut f\"ur Physik\\
Universit\"at Mainz\\
D-55099 Mainz, GERMANY}
\bigskip\bigskip
\end{raggedright}

\section{Introduction}
The single-flavour quark condensate \qq is a fundamental parameter of $\chi PT$, determining the relative size of mass and momentum terms in the expansion. Since it can not be predicted theoretically, its value must be determined experimentally, e.g.~by measuring the \pipi scattering lengths, whose values are predicted very precisely within the framework of $\chi PT$, assuming a big quark condensate \cite{gc2005}, or of generalised $\chi PT$, where the quark condensate is a free parameter \cite{mk1995}.   

The \dkpm decay is a very clean environment for the measurement of \pipi scattering lengths, since the two pions are the only hadrons and they are produced close to threshold. The only theoretical uncertainty enters through the constraint \cite{ba2001} between the scattering lengths \azd and \azz. In the \kthreepin decay a cusp-like structure can be observed at $\mnnsq=4\mpsq$, due to re-scattering from \kthreepic. The scattering lengths can be extracted from a fit of the \mnnsq distribution around the discontinuity.

\section{Experimental setup}
Simultaneous $K^+$ and $K^-$ beams were produced by 400 GeV energy protons from the CERN SPS, impinging on a beryllium target. The kaons were deflected in a front-end achromat in order to select the momentum band of $(60\pm 3)$ GeV/$c$ and focused at the beginning of the detector, about 200 m downstream. For the measurements presented here, the most important detector components are the magnet spectrometer, consisting of two drift chambers before and two after a dipole magnet and the quasi-homogeneous liquid krypton electromagnetic calorimeter. The momentum of the charged particles and the energy of the photons are measured with a relative uncertainty of 1\% at 20 GeV. A detailed description of the NA48/2 detector can be found in Ref.~\cite{rb2006}.  

\section{\boldmath \dkext}
\begin{figure}[htb]
  \begin{center}
    \begin{minipage}{3.1in}
      \begin{center}
        \includegraphics[width=3.1in,trim=50 320 50 50]{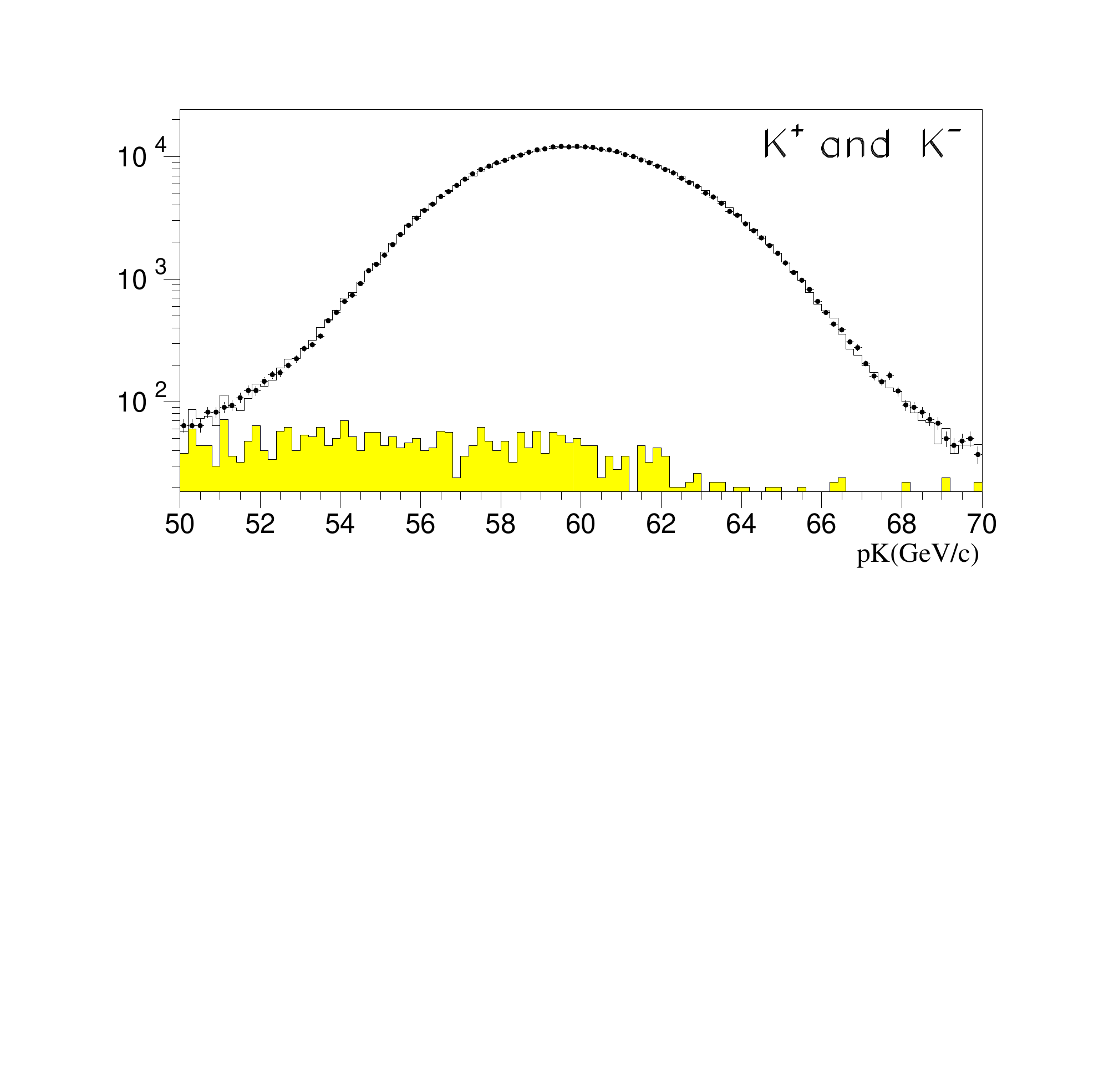}
      \end{center}
    \end{minipage}
    \hfill
    \begin{minipage}{2.5in}
      \begin{center}
        \includegraphics[width=2.5in,trim=50 50 20 0]{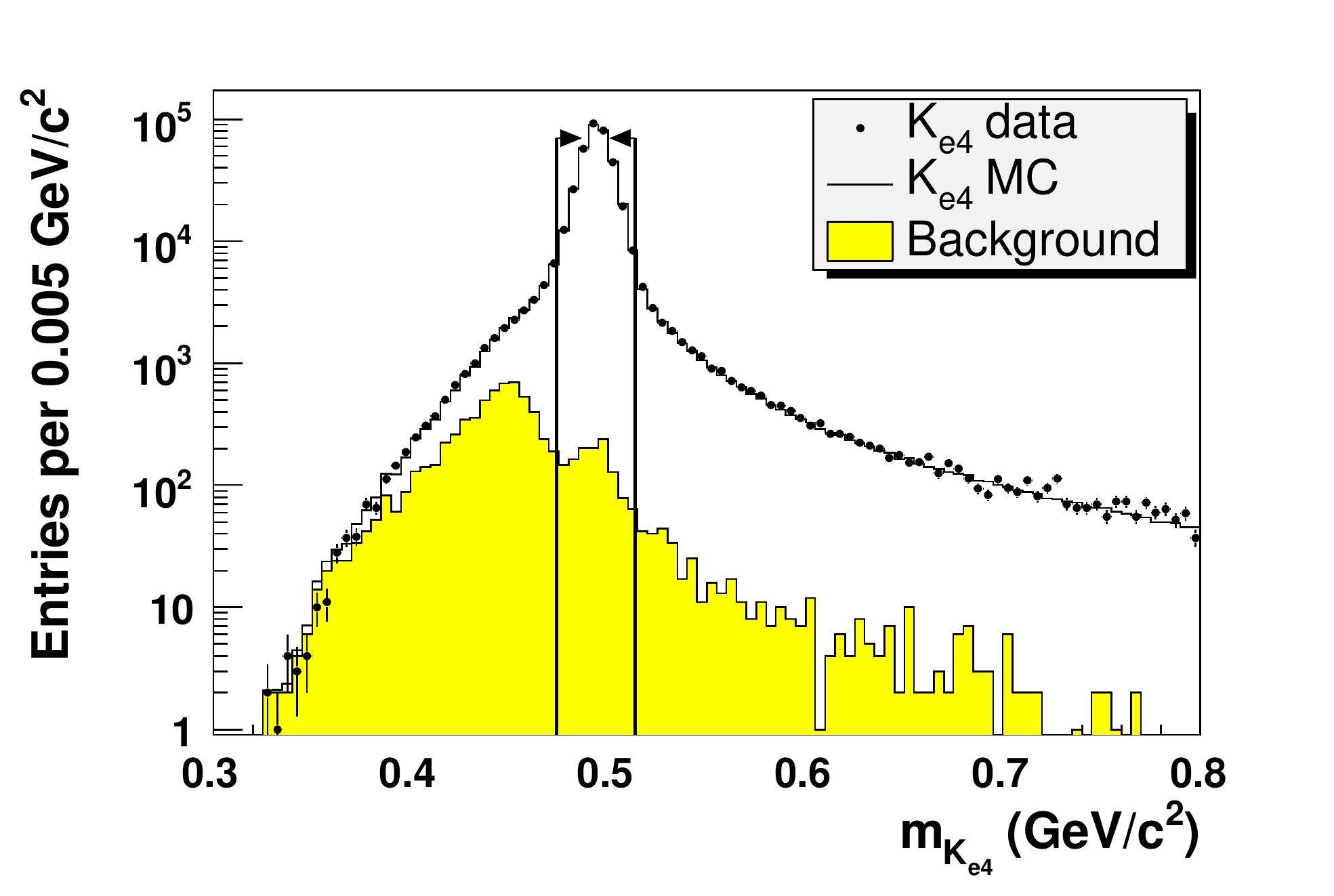}
      \end{center}
    \end{minipage}
  \end{center}
  \caption{\label{mke4pm} Kaon momentum (left) and mass (right) of the \dkpm events reconstructed with a quadratic or a linear equation, respectively. The data (crosses) are compared to signal MC (open histogram) plus background (yellow).}
\end{figure}

The \dkpm selection consisted of geometrical criteria, like the requirement of having three tracks within the detector acceptance and building a good vertex; particle identification requirements, based mainly on the different fraction of energy deposited by pions and electrons in the electromagnetic calorimeter; kinematical cuts for background rejection, like an elliptical cut in the ($p_T$,$M_{3\pi}$) plane centered at (0,$M_K$). In order to improve the pion rejection, the electron identification also included a Linear Discriminant Analysis combining the three quantities with the highest discriminating power. Two reconstruction strategies can be applied to the \dkpm events: either imposing the kaon mass and extracting the kaon momentum from a quadratic equation, or imposing the kaon momentum to be the mean beam momentum (60 GeV/$c$ along the beam axis) and extracting the kaon mass from a linear equation (see Fig.~\ref{mke4pm}).

Analysing part of the 2003 data, $3.7\times 10^5$ \dkpm events were selected with a background contamination below 1\%. The background level was estimated from data, using the so-called ``wrong sign'' events, i.e.~with the signature \wssel, that, at the present statistical level, can only be background, since the corresponding kaon decay violates the $\Delta S = \Delta Q$ rule and is therefore strongly suppressed \cite{pb1976}. The main background contributions are due to \kthreepic events with $\pi\to e\nu$ or a pion mis-identified as an electron. The background estimate from data was cross-checked using Monte Carlo simulation (MC).

\subsection{Form factors}
\begin{figure}[htb]
  \begin{center}
    \includegraphics[width=10cm,trim=0 20 10 20]{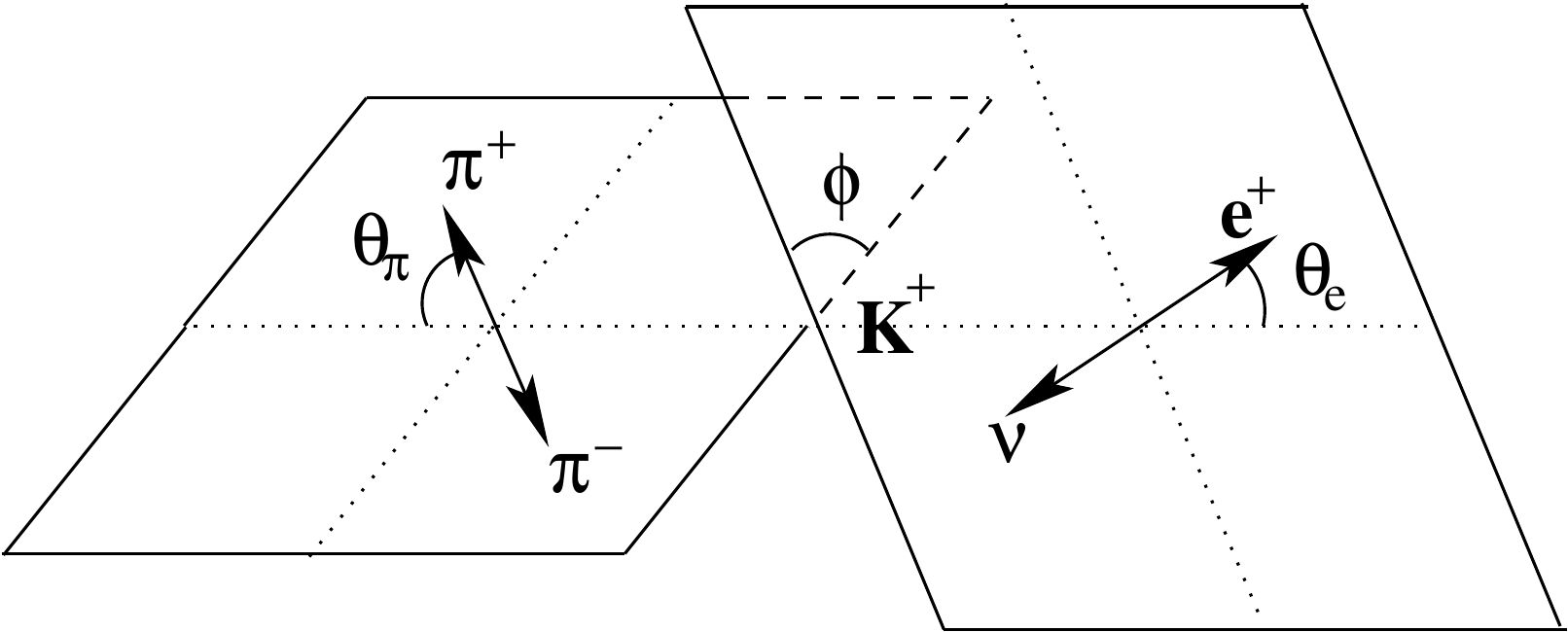}
  \end{center}
  \caption{\label{cmvar} Topology of the \dk decay.}
\end{figure} 

\begin{figure}[htb]
  \begin{center}
    \begin{minipage}{2.8in}
      \begin{center}
        \includegraphics[width=2.8in,trim=20 250 60 20]{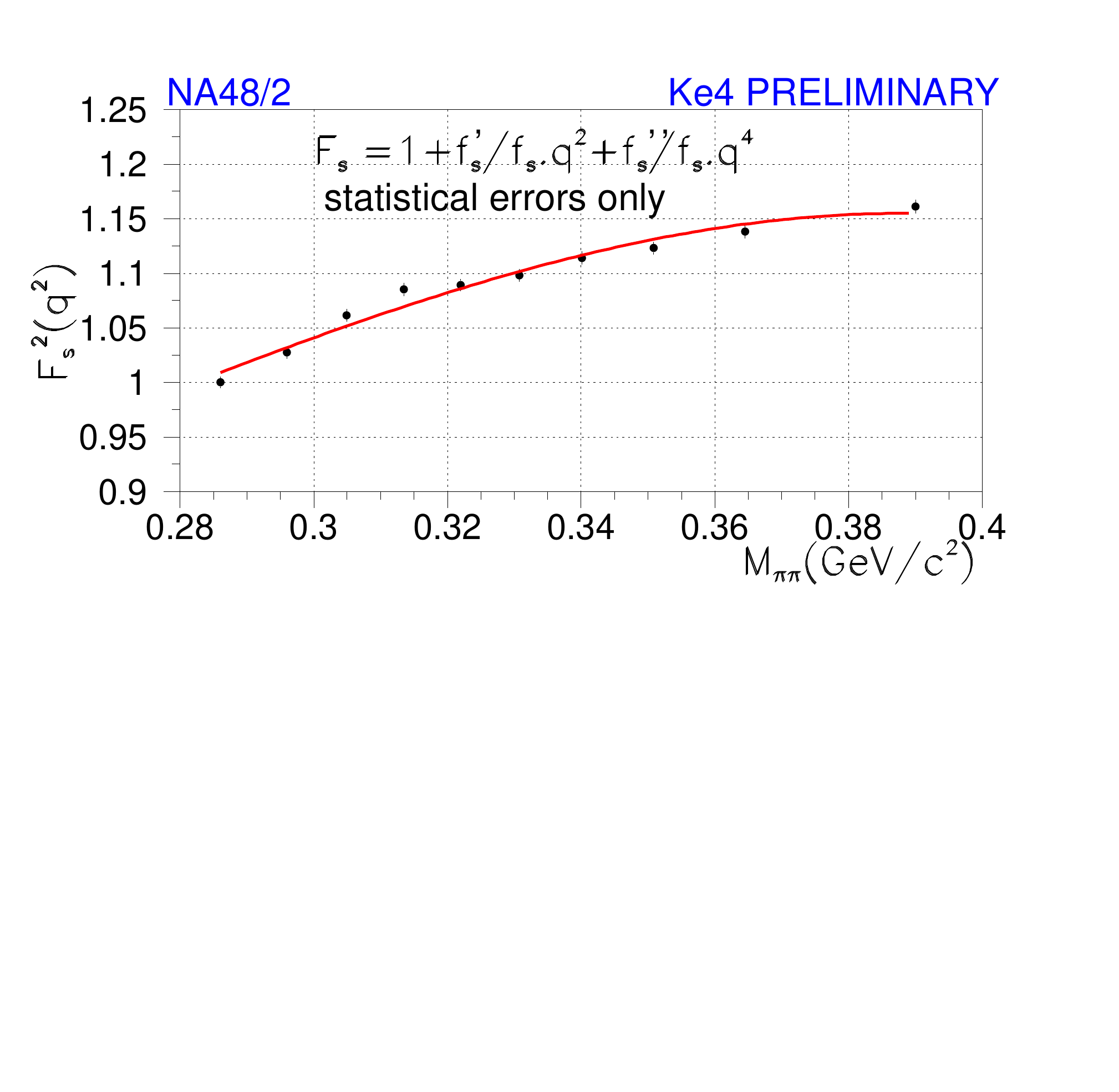}\\
        \includegraphics[width=2.8in,trim=20 270 60 20]{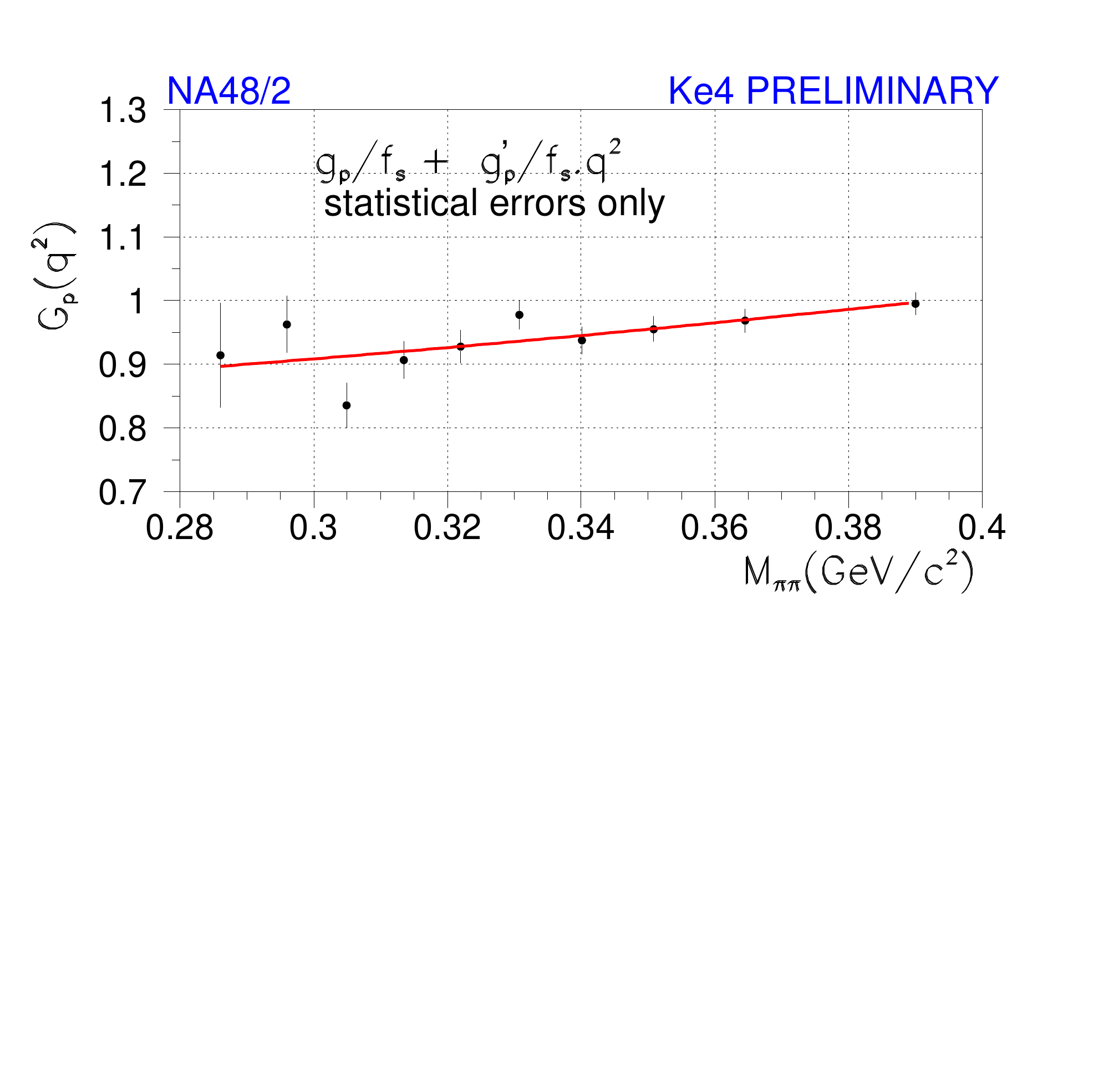}
      \end{center}
    \end{minipage}
    \hfill
    \begin{minipage}{2.8in}
      \begin{center}
        \includegraphics[width=2.8in,trim=50 250 40 20]{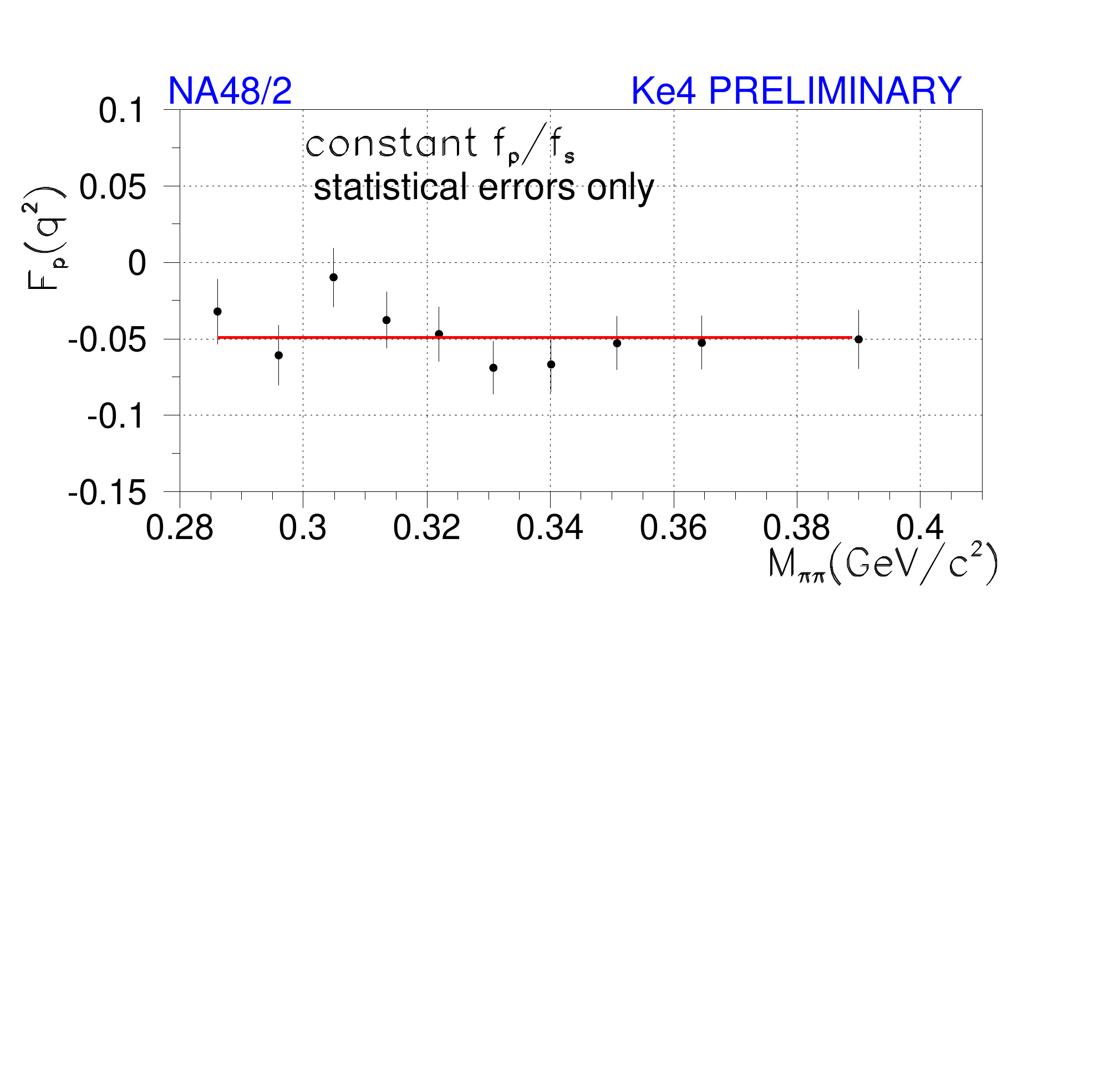}\\
        \includegraphics[width=2.8in,trim=50 270 40 20]{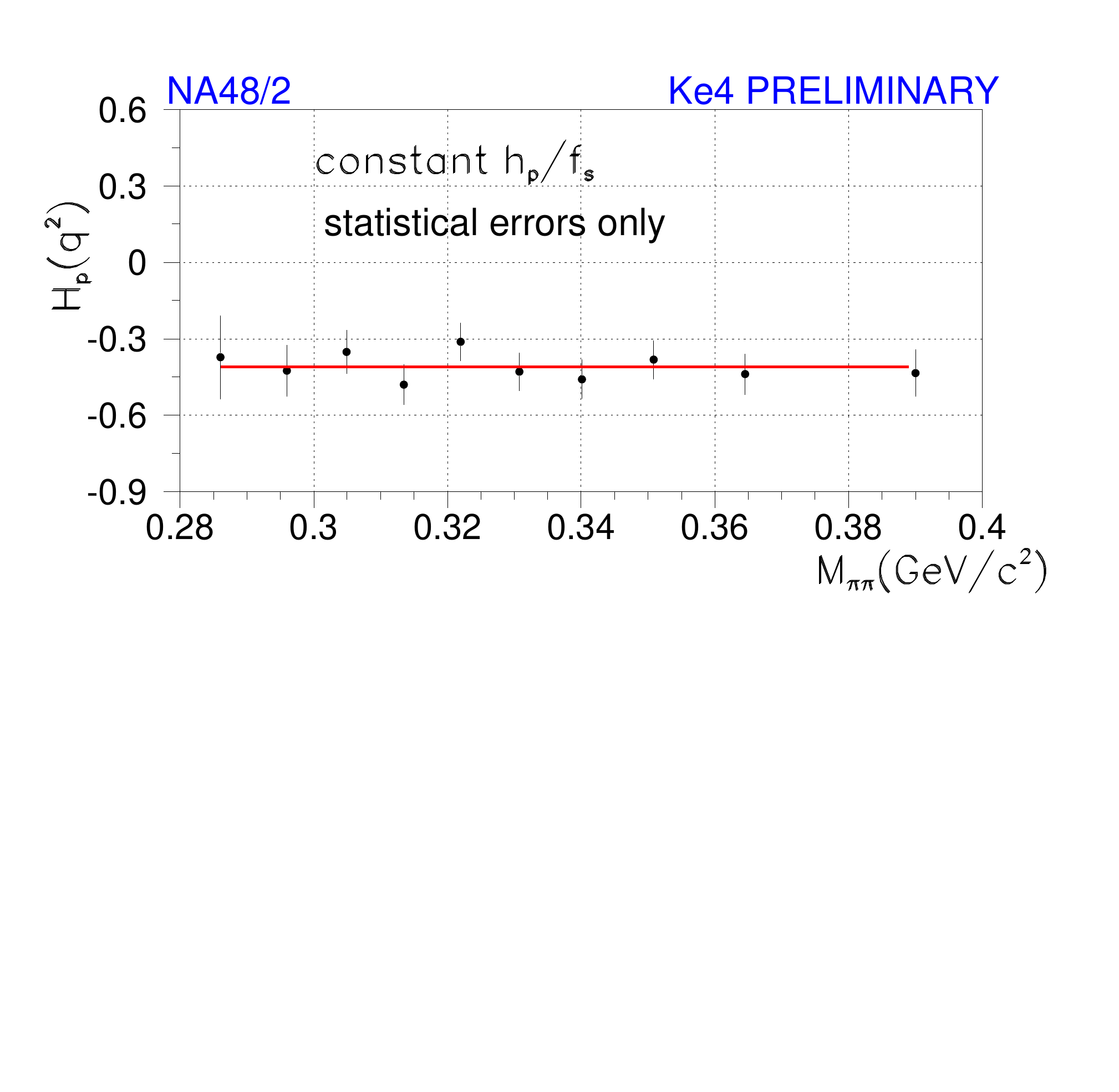}
      \end{center}
    \end{minipage}
  \end{center}
  \caption{\label{result} $F$, $G$ and $H$ dependence on $M_{\pipi}$. The points represent the results of the first-step fits, the lines are fitted in the second step.}
\end{figure}

\begin{figure}[htb]
  \begin{center}
    \includegraphics[width=11cm,trim=25 275 60 45]{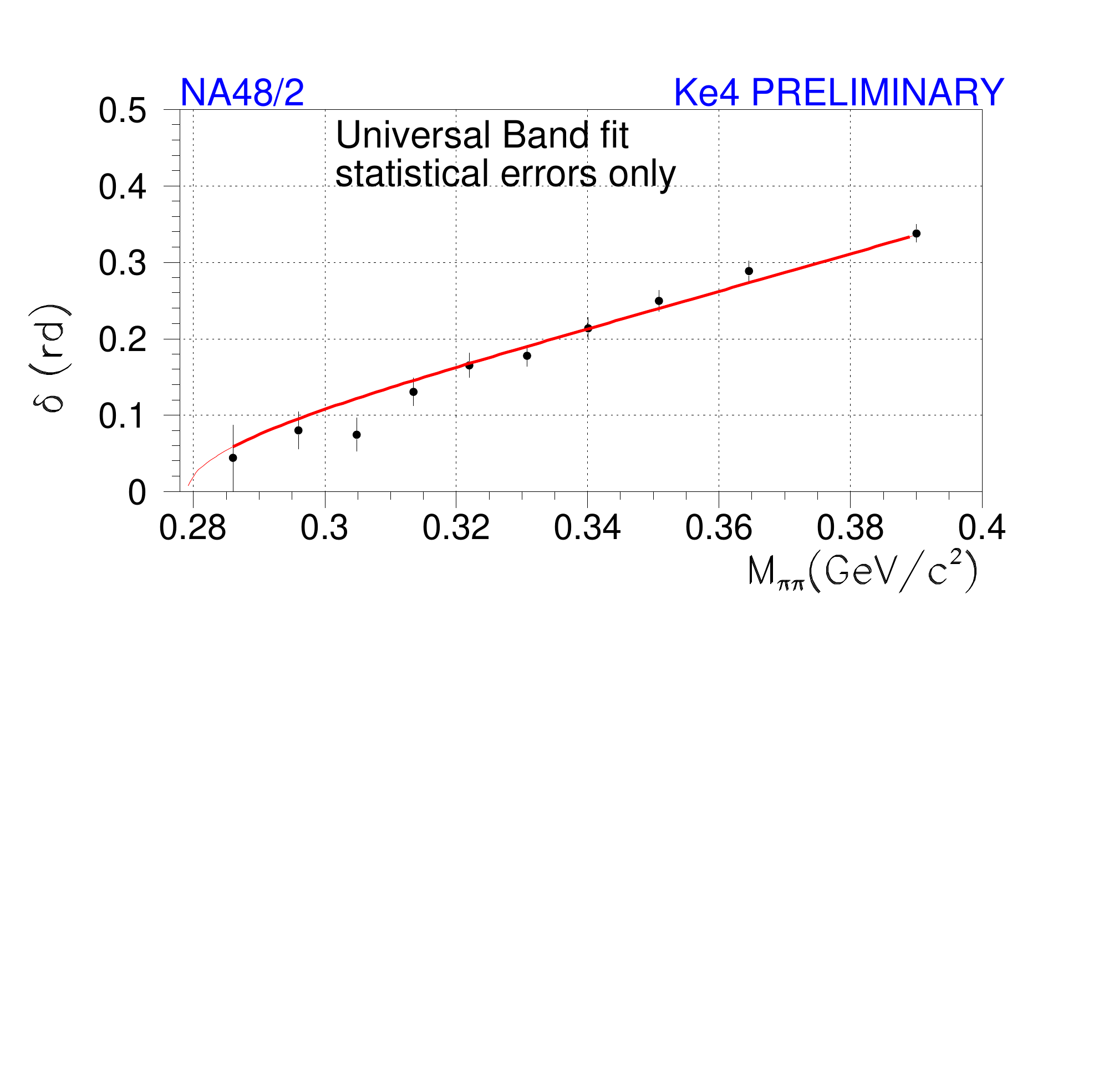}
  \end{center}
  \caption{\label{delta} $\delta = \delta_0^0-\delta_1^1$ distribution as a function of $M_{\pipi}$. The points represent the results of the first-step fits, the line is fitted in the second step.}
\end{figure}

The form factors of the \dkpm decay are parametrised as a function of five kinematic variables \cite{nc1965} (see Fig.~\ref{cmvar}): the invariant masses $M_{\pipi}$ and $M_{e\nu}$ and the angles $\theta_{\pi}$, $\theta_{e}$ and $\phi$.
The matrix element
\begin{displaymath}
  T=\frac{G_F}{\sqrt{2}}V_{us}^*\bar{u}(p_{\nu})\gamma_{\mu}(1-\gamma_5)v(p_e)(V^{\mu}-A^{\mu})
\end{displaymath}
contains a hadronic part, that can be described using two axial ($F$ and $G$) and one vector ($H$) form factors \cite{jb1995}. After expanding them into partial waves and into a Taylor series in $q^2=M_{\pipi}^2/4\mpsq-1$, the following parametrisation was used to determine the form factors from the experimental data \cite{ap1968,ga1999}:
\begin{eqnarray*}
  F &=& (f_s+f'_s q^2+f''_s q^4) e^{i\delta_0^0(q^2)}+f_p \cos{\theta_{\pi}} e^{i\delta_1^1(q^2)}\\
  G &=& (g_p + g'_p q^2) e^{i\delta_1^1(q^2)}\\
  H &=& h_p e^{i\delta_1^1(q^2)}.
\end{eqnarray*}
In a first step, ten independent five-parameter fits were performed for each bin in $M_{\pipi}$, comparing data and MC in four-dimensional histograms in $M_{e\nu}$, $\cos\theta_{\pi}$, $\cos\theta_e$ and $\phi$, with 1500 equal population bins each. The second step consisted in a fit of the distributions in $M_{\pipi}$ (see Figs.~\ref{result},\ref{delta}), to extract the (constant) form factor parameters.

The polynomial expansion in $q^2$ was truncated according to the experimental sensitivity. The dependence on $M_{e\nu}$ and the $D$-wave were found to be negligible within the total uncertainty and the corresponding parameters were therefore set to zero. The $\delta = \delta_0^0-\delta_1^1$ distribution was fitted with a one-parameter function given by the numerical solution of the Roy equations \cite{ba2001}, in order to determine \azz, while \azd was constrained to lie on the centre of the universal band. The following preliminary result was obtained:
\renewcommand{\arraystretch}{1.2}
\begin{eqnarray*}
  f'_s/f_s &=& \phantom{-}0.169\pm 0.009_{stat}\pm 0.034_{syst}\\
  f''_s/f_s &=& -0.091\pm 0.009_{stat}\pm 0.031_{syst}\\
  f_p/f_s &=& -0.047\pm 0.006_{stat}\pm 0.008_{syst}\\
  g_p/f_s &=& \phantom{-}0.891\pm 0.019_{stat}\pm 0.020_{syst}\\
  g'_p/f_s &=& \phantom{-}0.111\pm 0.031_{stat}\pm 0.032_{syst}\\
  h_p/f_s &=& -0.411\pm 0.027_{stat}\pm 0.038_{syst}\\
  a_0^0 &=& \phantom{-}0.256\pm 0.008_{stat}\pm 0.007_{syst}\pm 0.018_{theor},
\end{eqnarray*}
where the systematic uncertainty was determined by comparing two independent analyses and taking into account the effect of reconstruction method, acceptance, fit method, uncertainty on background estimate, electron-ID efficiency, radiative corrections and bias due to the neglected $M_{e\nu}$ dependence. The form factors are measured relative to $f_s$, which is related to the decay rate. The obtained value for \azz is compatible with the $\chi PT$ prediction $\azz=0.220\pm 0.005$ \cite{gc2001} and with previous measurements \cite{lr1977,sp2003}. 

\section{\boldmath \dkextnn}
\begin{figure}[htb]
  \begin{center}
    \includegraphics[width=9cm,trim=25 40 0 0]{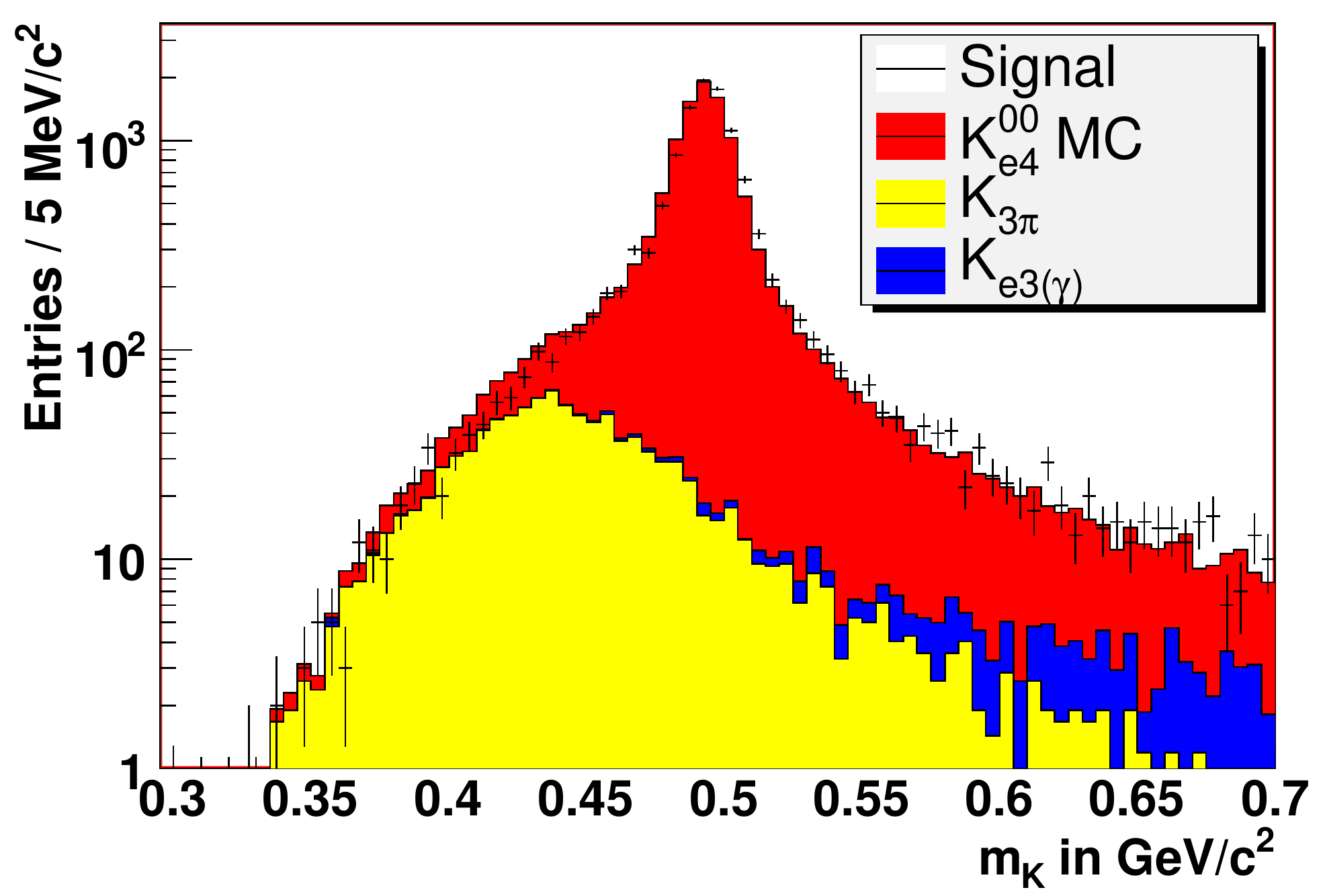}
  \end{center}
  \caption{\label{ke400} Invariant mass distribution in logarithmic scale of the \dknn events selected from the 2003 data (crosses) compared to the signal MC (red) plus physical (yellow) and accidental (blue) background.}
\end{figure}

About 10,000 \dknn events were selected from the 2003 data and about 30,000 from the 2004 data with a background contamination of 3\% and 2\%, respectively. The selection criteria were similar to the ones used for the \dkpm events, apart from the requirement of containing one track and 4 photons compatible with two $\pi^0$s at the same vertex. The electron identification was based on the fraction of energy deposited in the electromagnetic calorimeter and on the width of the corresponding shower. The background level was estimated from data by reversing some of the selection criteria and was found to be mainly due to \kthreepin events with a pion mis-identified as an electron (see Fig.~\ref{ke400}).

The branching fraction was measured, as a preliminary result from the 2003 data only, normalised to \kthreepin:
\begin{eqnarray*}
  BR(\dknn)&=&(2.587 \pm 0.026_{stat} \pm 0.019_{syst} \pm 0.029_{ext})\times 10^{-5},
\end{eqnarray*}
where the systematic uncertainty takes into account the effect of acceptance, trigger efficiency and energy measurement of the calorimeter, while the external uncertainty is due to the uncertainty on the \kthreepin branching fraction. This result is about eight times more precise than the best previous measurement \cite{ss2004}.

\begin{figure}[htb]
  \begin{center}
    \includegraphics[width=10cm,trim=25 40 0 0]{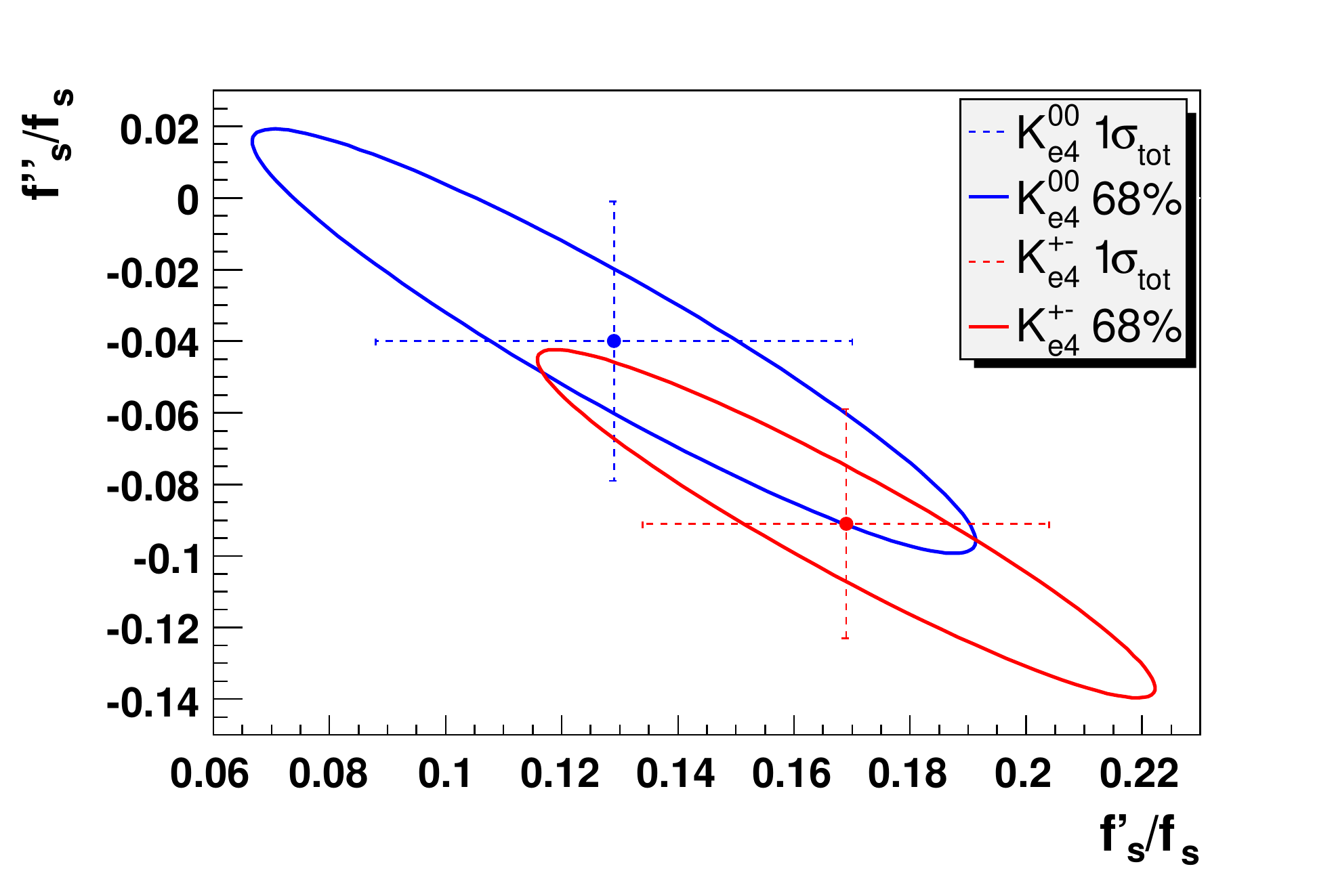}
  \end{center}
  \caption{\label{ellfs} Comparison of the $f'_s/f_s$ and $f''_s/f_s$ measurements in \dkpm and \dknn.}
\end{figure}

For the form factors the same formalism is used as in \dkpm, but, due to the symmetry of the $\pi^0\pi^0$ system, the $P$-wave is missing and only two parameters are left: $f'_s/f_s$ and $f''_s/f_s$. Using the full data sample, the following preliminary result was obtained:
\begin{eqnarray*}
  f'_s/f_s &=& \phantom{-}0.129\pm 0.036_{stat}\pm 0.020_{syst}\\
  f''_s/f_s &=& -0.040\pm 0.034_{stat}\pm 0.020_{syst},
\end{eqnarray*} 
which is compatible with the \dkpm result (see Fig.~\ref{ellfs}).

\section{\boldmath \kthreepin}
\begin{figure}[htb]
  \begin{center}
    \begin{minipage}{2.8in}
      \begin{center}
        \includegraphics[width=2.8in,trim=20 40 60 70]{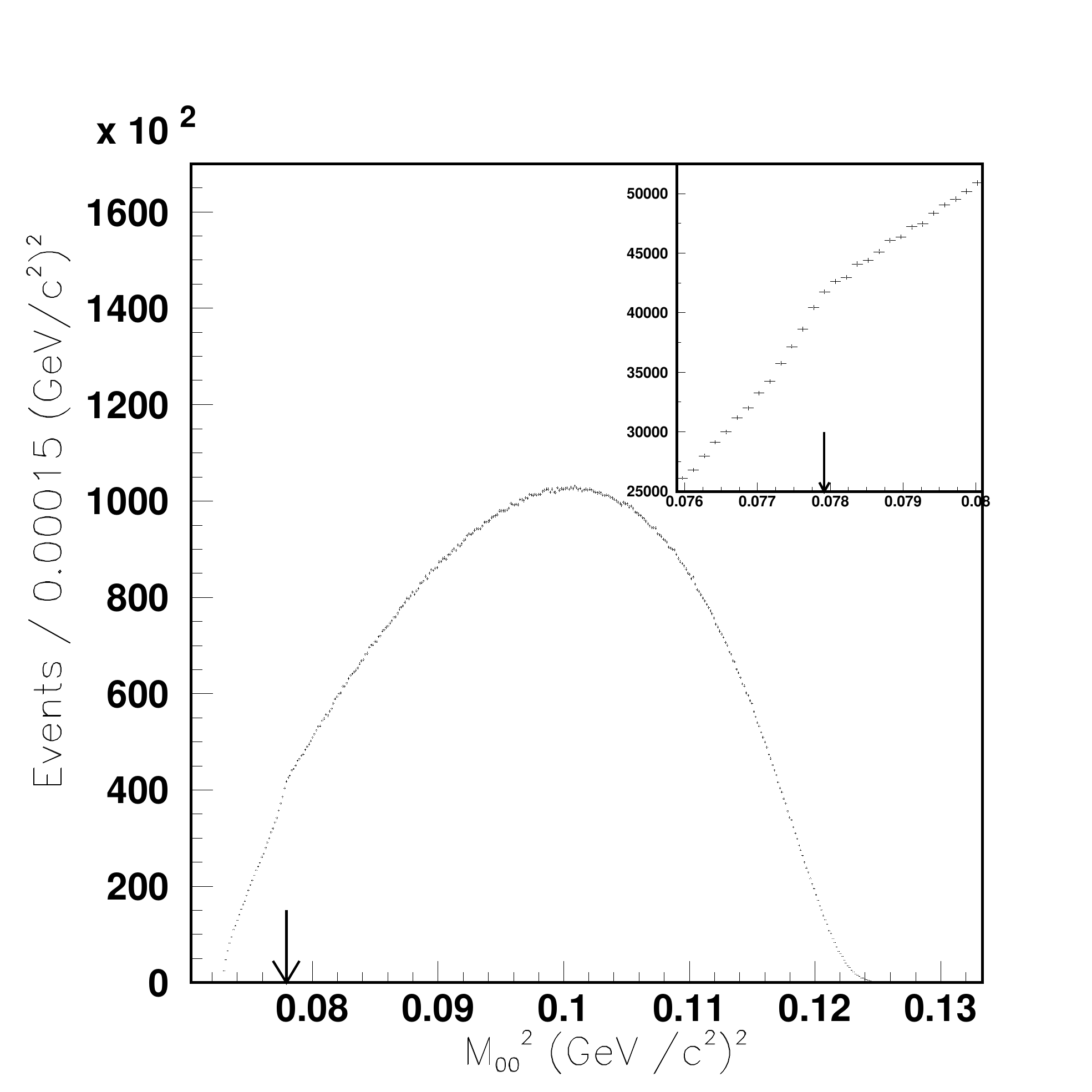}
      \end{center}
    \end{minipage}
    \hfill
    \begin{minipage}{2.8in}
      \begin{center}
        \includegraphics[width=2.8in,trim=20 40 60 70]{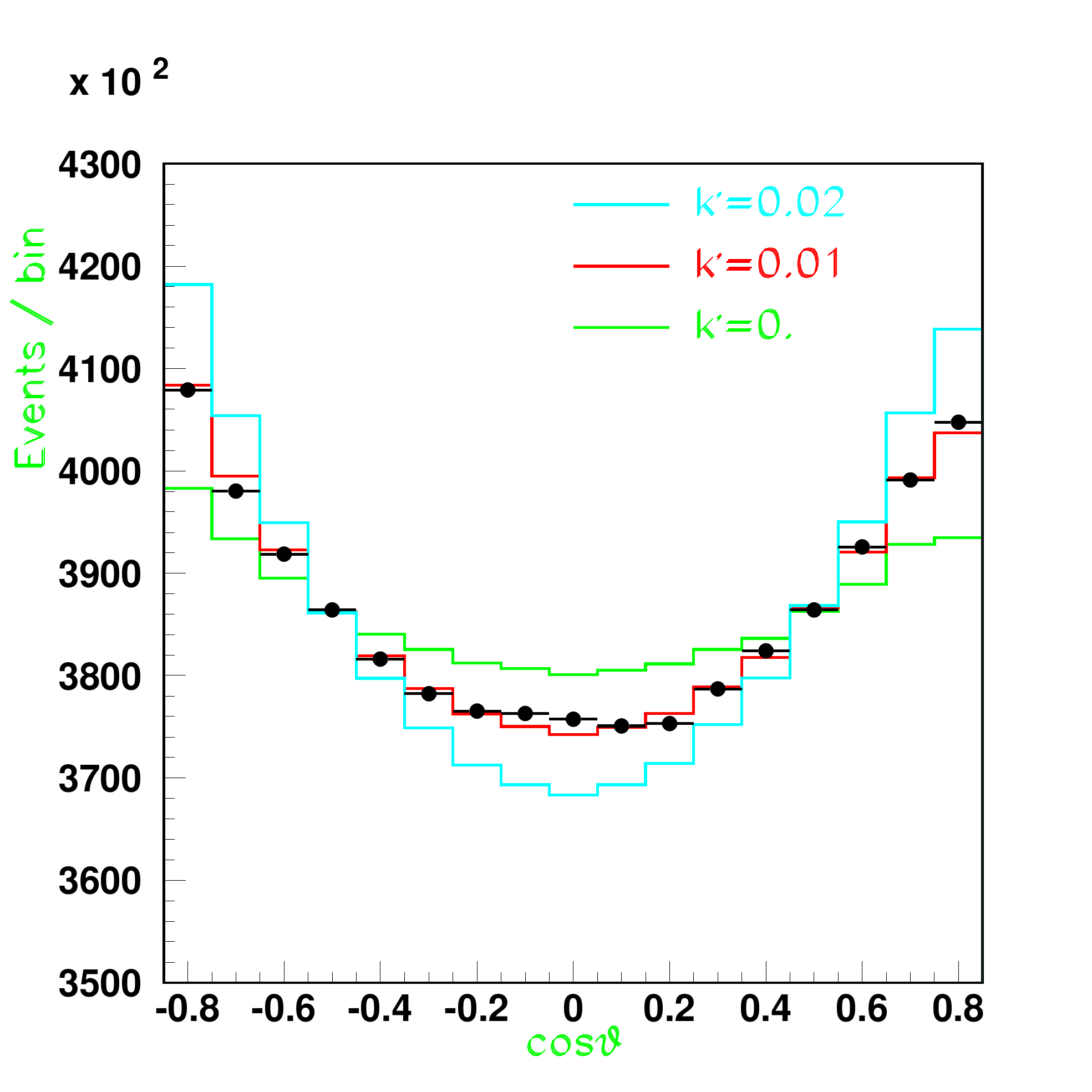}
      \end{center}
    \end{minipage}
  \end{center}
  \caption{\label{cusp} Left: \mnnsq of the selection \kthreepin data events. The arrow indicates the position of the cusp. Right: angle between the $\pi^{\pm}$ and the $\pi^0$ in the $\pi^0\pi^0$ centre of mass system. The points represent the data, the three curves, the MC distribution for different values of $k'$}
\end{figure}

From 2003 data, about 23 million \kthreepin events were selected, with negligible background. The squared invariant mass of the $\pi^0\pi^0$ system (\mnnsq) was computed imposing the mean vertex of the $\pi^0$s, in order to improve its resolution close to threshold. At $\mnnsq=4\mpsq$, the distribution shows evidence for a cusp-like structure (see Fig.~\ref{cusp}, left) due to \pipi re-scattering. 

Fitting the distribution with the theoretical model presented in Ref.~\cite{nc2005} and using the unperturbed matrix element
\begin{center}
  $M_0 = A_0 (1 + \frac{1}{2}g_0u+\frac{1}{2}h'u^2+\frac{1}{2}k'v^2),$
\end{center}
the following result was obtained \cite{rb2005}, assuming $k'=0$ \cite{PDG}:
\begin{eqnarray*}
  g_0 &=& \phantom{-}0.645\pm 0.004_{stat}\pm 0.009_{syst}\\
  h' &=& -0.047\pm 0.012_{stat}\pm 0.011_{syst}\\
  a_2 &=& -0.041\pm 0.022_{stat}\pm 0.014_{syst}\\
  a_0-a_2 &=& \phantom{-}0.268\pm 0.010_{stat}\pm 0.004_{syst}\pm 0.013_{theor},
\end{eqnarray*}
where the $a_0-a_2$ measurement is dominated by the uncertainty on the theoretical model.

In a further analysis, the value of $k'$ was obtained from a fit above the cusp in the plane $\cos{\theta}$ vs \mnnsq, where $\theta$ is the angle between the $\pi^+$ and the $\pi^0$ in the $\pi^0\pi^0$ centre of mass system. Evidence was found for a non-zero value of $k'$ (see Fig.~\ref{cusp}, right):
\begin{displaymath}
  k'=0.0097\pm 0.0003_{stat}\pm 0.0008_{syst},
\end{displaymath}
where the systematic uncertainty takes into account the effect of acceptance and trigger efficiency.
Reweighting the MC with the obtained value of $k'$, the standard fit of the \mnnsq distribution with the Cabibbo-Isidori model was performed to obtain the cusp parameters, that were found to be compatible with the published values.



\end{document}

%% file: econfmacros.tex
\def\beq{\begin{equation}}
\def\eeq#1{\label{#1}\end{equation}}
\def\eeqn{\end{equation}}

\def\beqa{\begin{eqnarray}}
\def\eeqa#1{\label{#1}\end{eqnarray}}
\def\eeqan{\end{eqnarray}}

\let\bar=\overbar

\def\Dslash{\not{\hbox{\kern-4pt $D$}}}
\def\dslash{\not{\hbox{\kern-2pt $\del$}}}

\def\msb{{\bar{\ssstyle M \kern -1pt S}}}

